\documentclass[twocolumn,showpacs,preprintnumbers,amsmath,amssymb]{revtex4}
\usepackage{graphicx}
\usepackage{dcolumn}
\usepackage{bm}

\begin{document}

\preprint{APS/123-QED}
\title{Superradiant Rayleigh scattering in a ring cavity}

\author{S. Slama}
\author{S. Bux}
\author{G. Krenz}
\author{C. Zimmermann}
\author{Ph.W. Courteille}
\affiliation{Physikalisches Institut, Eberhard-Karls-Universit\"at T\"ubingen,
\\Auf der Morgenstelle 14, D-72076 T\"ubingen, Germany}

\date{\today}

\begin{abstract}
Collective interaction of light with an atomic gas can give rise to superradiant instabilities. We experimentally study the sudden build-up of a reverse light field in a laser-driven high-finesse ring cavity filled with ultracold thermal or condensed atoms. While superradiant Rayleigh scattering from atomic clouds is normally only observed at very low temperatures (i.e.~well below $1~\mu$K), the presence of the ring cavity enhances cooperativity and allows for superradiance with thermal clouds as hot as several $10~\mu$K. A characterization of the superradiance at various temperatures and cooperativity parameters allows us to link it to the collective atomic recoil laser.
\end{abstract}


\maketitle

Cold atoms are most interesting candidates for studying collective instabilities. On one hand, they can easily be prepared and controlled in large numbers accessible to thermodynamic descriptions. On the other hand, individual atoms follow the rules of quantum mechanics and quantum statistics at ultra-low temperatures, which predestinates them for studies of quantum synchronization effects \cite{Zhirov06}.

Collective instabilities in clouds of cold and ultra-cold atoms driven by light have recently been observed in various situations \cite{Inouye99,Kruse03b,Nagorny03b}, the most prominent of which are superradiant Rayleigh scattering (SRS) \cite{Inouye99} and collective atomic recoil lasing (CARL) \cite{Kruse03b}. The signature of CARL is the sudden build-up of a probe field oriented reversely to a strong pump interacting with an atomic gas together with a bunching of the atoms \cite{Bonifacio95}. The underlying runaway amplification mechanism is particularly strong, if the reverse probe field is recycled by a ring cavity. In previous papers, we have discussed the relationship between CARL, recoil-induced resonances \cite{Kruse03b} and Kuramoto-type self-synchronization \cite{Cube04}. But CARL also has a close analogy with SRS, since they both share the same gain mechanism \cite{Piovella01}.

SRS has first been observed by shining a short laser pulse into a Bose-Einstein condensate (BEC) \cite{Inouye99}. The scattering atoms collectively form motional sidemodes that are coupled out of the condensate, while at the same time a burst of light is emitted into the long axis of the condensate. The phenomenon has initially been explained as a four-wave mixing process between optical and matter wave modes with bosonic stimulation through the macroscopically populated final atomic momentum state. Although it was realized that bosonic quantum degeneracy was not essential for SRS, the bosonic enhancement picture, which seemed corroborated by the fact that SRS was (at first) not observable for thermal atomic clouds, instigated a controversy about the role of quantum statistics. Theoretical work \cite{Moore01} and the very recent experimental observation of SRS with thermal atoms \cite{Yoshikawa05} showed however that the gain mechanism is independent on quantum statistics. What counts is not the quantum nature of the particles, but their \textit{cooperative} behavior.

The difficulty with thermal atomic ensembles is that the coherence time ruling the observability of SRS is Doppler-limited \cite{Yoshikawa05}. It is therefore surprising that collective gain can be seen for CARL with thermal atoms as hot as a few $100~\mu$K \cite{Kruse03b}. In the present work we demonstrate that SRS is also possible at high temperature, and we show how the presence of an optical high-finesse ring cavity preserves coherence and cancels out the deleterious effects of thermal motion. In fact, the superradiant gain process itself causes a rapid diffusion in momentum space leading to decoherence, unless the cavity restricts the density-of-states available to the scattered light mode. The same mechanism works for thermal diffusion. The spectrum of the scattered light, which is Doppler-broadened by thermal atomic motion, is filtered out by the cavity. The experimental signatures we present thus demonstrate the common root of CARL and SRS.

\begin{figure}[ht]
		\centerline{\scalebox{0.46}{\includegraphics{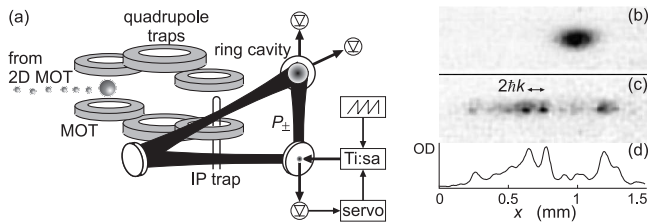}}}\caption{
			(a) Schematic view of the experimental setup. A two-dimensional MOT (2D MOT) feeds a MOT in the main chamber. From here the
			cloud is transferred adiabatically in several intermediate steps into a Ioffe-Pritchard (IP) type magnetic trap overlapping
			with the ring cavity mode volume. A Ti:sapphire laser resonantly pumps the cavity mode $P_+$. Both cavity modes
			$P_\pm$ are observed via the light fields leaking out through one of the cavity mirrors. The atomic cloud can be visualized by
			absorption imaging. Typical images of a condensate cloud at $T=0.5T_c$ having and not having interacted with the cavity are
			shown in (b) and (c), respectively. The images are recorded after $10~$ms of free expansion. Curve (d) shows the vertically
			integrated optical density (OD) of image (c).}
		\label{Fig1}
	\end{figure}

Our experiment represents the first study of a BEC stored in a macroscopic optical cavity. We produce a cloud of ultracold $^{87}$Rb in a magnetic trap and then move the cloud into the mode volume of a high-finesse optical ring cavity. The scheme of our experiment is shown in Fig.~\ref{Fig1}(a). We load the atoms from a two-dimensional magneto-optical trap (MOT) into a standard MOT located in an ultrahigh vacuum chamber. From here the atoms are transferred into a magnetic quadrupole potential operated with the same coils as the MOT. The potential is then compressed and the atoms are transferred via a second into a third quadrupole trap. With two pairs of thin wires separated by $1~$mm and running parallel to the symmetry axis of the third quadrupole trap a Ioffe-Pritchard type potential is created \cite{Silber05}. The atoms can easily be shifted up and down along the Ioffe wires.

In the Ioffe-Pritchard trap the Rb cloud is cooled by forced evaporation: A microwave frequency resonantly tuned to the ground state hyperfine structure couples the trapped Zeeman state $|2,2\rangle$ and the untrapped $|1,1\rangle$. After $15~$s of down-ramping the microwave, we reach the threshold to quantum degeneracy at $T_{\mathrm{c}}=800~$nK with about $N=5\times10^5$ atoms. Typical trap frequencies at the end of the evaporation ramp are $\omega_x\approx\omega_y\simeq 2\pi\times200~$Hz and $\omega_z\simeq 2\pi\times50~$Hz, obtained at a bias field of $2~$G. Cooling down further yields almost pure condensates of $2\times10^5$ Rb atoms. The long axis of the cigar-shaped condensate is parallel to the ring cavity axis.

The ring cavity is very similar to the one used in Refs.~\cite{Kruse03,Kruse03b,Cube04}. It consists of one plane and two curved mirrors. It is $8.5$~cm long corresponding to a free spectral range of $\delta_{fsr}=3.5~$GHz and has a beam waist of $w_0=107~\mu$m. One mode is continuously pumped by a titanium-sapphire laser. The laser can be stabilized to this mode using the Pound-Drever-Hall method. The ring cavity can be operated in two ways depending on the polarization of the incoupled light. For $p$-polarized light a finesse of 87000 is determined from a measured intensity decay time of $\tau\approx3.8~\mu$s. For $s$-polarized light the finesse is 6400.

We measure the intracavity light powers in the pump mode, $P_+$, and the reverse mode, $P_-$, via the fields leaking through one of the cavity mirrors [see Fig.~\ref{Fig1}(a)]. In order to prevent atom losses due to light scattering, we switch off the pump laser while the atoms are driven into the resonator. As soon as the magnetically trapped atoms are located within the mode of the ring cavity, we switch on the Ti:sapphire laser tuned between $0.5$ and $2~$nm red from the $D_1$-line at $794.8~$nm and ramp the laser frequency across one of the cavity resonances. When the laser crosses a cavity resonance, the fast branch of the Pound-Drever-Hall servo (controlling an acousto-optic modulator) quickly pulls the laser frequency to the center of the cavity resonance and tightly locks its phase. When the slow branch (which controls a piezo transducer mounted to the laser cavity) is interrupted the capture range of the servo is limited to a few MHz. Hence, when the ramp goes beyond this capture range, the servo looses its grip and the laser leaves the cavity resonance again. The build-up time for the cavity mode $\alpha_+$ is limited by the bandwidth of the locking servo to about $\tau_{bw}=20~\mu$s, which is a bit longer than the cavity decay time.

Another observable is the atomic density distribution measured by time-of-flight absorption imaging. A typical image of an expanded condensate having interacted with the cavity fields is shown in Fig.~\ref{Fig1}(c). The intensity of the various Bragg diffraction orders tell the population of the momentum states.

\begin{figure}[ht]
		\centerline{\scalebox{0.49}{\includegraphics{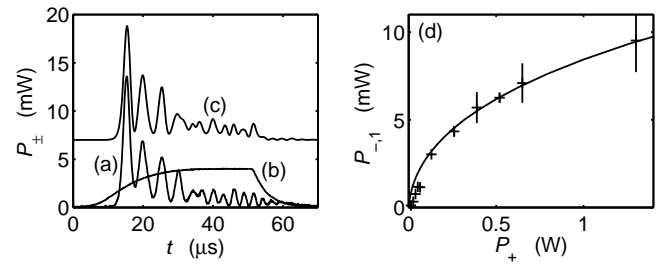}}}\caption{
			(a) Measured time-evolution of the reverse power $P_-$. The pump laser power is $P_+=4~$W. The atom number is $N=1.5\times10^6$
			and the laser wavelength is $\lambda=797.3~$nm. Curve (b) marks the time-evolution of the recorded pump laser power scaled down
			by 1000. Curve (c) shows (offset by $7~$mW) a numerical simulation of the reverse power using the above parameters (see text).
			To account for the finite switch-on time of the pump laser power, its experimentally recorded time-evolution is plugged into
			the simulations, where we assume that the pump laser frequency is fixed and resonant to a cavity mode. 
			(d) Measured and calculated (solid line) height $P_{-,1}$ of the first superradiant peak as a function of pump power $P_+$.
			Here $N=2.4\times10^6$ and $\lambda=796.1~$nm.}
		\label{Fig2}
	\end{figure}

Fig.~\ref{Fig2}(a,b) shows a typical recorded time evolution of the powers $P_{\pm}$. It illustrates the central phenomenon studied in this paper. The pumped mode fills up with light on the time scale $\tau_{bw}$. In constrast, the power build-up in the reverse mode is delayed and suddenly increases exponentially, until it reaches a level of 0.1 to 1\% of the pumped mode power. This superradiant burst of light is followed by several revivals and decays. The delay, the height of the first superradiant peak and the subsequent evolution depend on certain experimentally tunable parameters. Those are the atom number $N$, the pump power $P_+$, the temperature $T$ of the atom cloud, and the coupling strength $U_0=g^2/\Delta_a$, where $g=\sqrt{3\Gamma\delta_{fsr}}/kw_0\simeq2\pi\times84~$kHz is the one-photon Rabi frequency in our cavity, $\Delta_a$ the laser detuning from the atomic resonance, whose decay width is $\Gamma=2\pi\times6~$MHz, and $k=2\pi/\lambda$. The rise-time of the superradiant burst is about $1~\mu$s, which is considerably faster than the cavity mode build-up time. The $2~\mu$K cold atoms are not Bose-condensed. Our experiments do not show a qualitative difference in the reverse mode behavior below $T_c$.

The coupled dynamics is governed by the interdependence of only a few characteristic quantities \cite{Bonifacio95,Bonifacio97,Piovella01b}. Those are the cavity decay rate $\kappa=(2\tau)^{-1}\approx2\pi\times20~$kHz, the recoil shift $\omega_r=2\hbar k^2/m\approx 2\pi\times14~$kHz, the Doppler width of the atomic velocity distribution $\sigma_v=2k\sqrt{k_BT/m}$, and the (small signal) collective gain $G$. The gain describes the exponential increase of the number of recoiling atoms and of the scattered light intensity due to cooperation. For conventional SRS, i.e.~in the absence of a cavity, the relevant lifetime of the optical fields is limited by the size $L$ of the condensate along the axis into which the light is scattered, $\tau_{sr}=(2\kappa_{sr})^{-1}=L/c$. The gain is given by $G=N\Omega^2g_{sr}^2/2\kappa_{sr}\Delta_a^2$, where $\Omega$ is the Rabi frequency generated by the incident laser beam and $g_{sr}=\sqrt{3\Gamma\kappa_{sr}}/kw_{sr}$ is the coupling strength of the effective cavity formed by the condensate to the atomic $D_1$ transition. $w_{sr}$ is the radial width of the condensate. Typically, the gain $G\simeq10^5~\text{s}^{-1}$ is much smaller than $\kappa_{sr}$ \cite{Inouye99}, i.e.~scattered photons leave the interaction region before the next photons are scattered. For collective gain to take place, the coherence time must be longer than the mean time delay between subsequent scattering events, which is only possible if the coherence is stored as a Raman coherence between momentum states of the individual atoms before and after scattering. 

In contrast, if the decay of the scattered light mode is slowed down by a cavity, the cavity plays a major role in storing the coherence. The gain is then estimated by the same formula as above, but replacing $g_{sr}$ by $g$ and $\kappa_{sr}$ by $\kappa$. Furthermore, since the pump and the reverse beam now share the same mode volume, we may set $\Omega_i=\sqrt{n}g$. We obtain a gain, $G=nNU_0^2/2\kappa$, which can experimentally be tuned between $G\simeq10^3$~-~$10^9~\text{s}^{-1}$.

The various regimes in which CARL and SRS may occur are characterized by the size of the \textit{collective gain bandwidth} $\Delta\omega_G$ (which is the spectral regime in which the gain is larger than half its maximum value) \cite{Bonifacio95} as compared to the decay width $\kappa$ (or $\kappa_{sr}$) and the recoil frequency $\omega_r$. In particular, one may distinguish two situations called the \textit{superradiant} limit, where $\Delta\omega_G\ll\kappa$ and the \textit{good-cavity} limit for which $\Delta\omega_G\gg\kappa$ \cite{Bonifacio95}. Due to the fact that the field decay rate in our ring cavity is 6 to 7 \textit{orders of magnitude smaller} than in the superradiance experiments, both regimes are accessible to our experiment by operating the cavity at either high or low finesse and by appropriately tuning the gain via the parameters $P_+$, $N$, and $\lambda$. At a given cavity decay width the superradiant regime is attained for weak collective coupling, i.e.~small atom numbers and low pump powers, while the good-cavity limit is realized for large atom numbers and high powers \cite{Bonifacio97}. 

Fig.~\ref{Fig2}(d) shows the dependence of the height of the first superradiant peak as a function of the pump laser power in the good-cavity limit, i.e.~for the high finesse case. The theoretical curves in this and all other figures are calculated based on the CARL model already used in Ref.~\cite{Kruse03b,Bonifacio95,Bonifacio97,Gangl00}, which treats all degrees of freedom \textit{classically}. It also adiabatically eliminates the electronically excited states, which is a good approximation far from resonance \cite{Note1}. The simulations are done with 100 atoms, each atom representing $N/100$ atoms. Finally, we tested that the classical model coincides with a generalized description treating the atomic motion as quantized to a good approximation \cite{Piovella01b}.

\begin{figure}[ht]
		\centerline{\scalebox{0.48}{\includegraphics{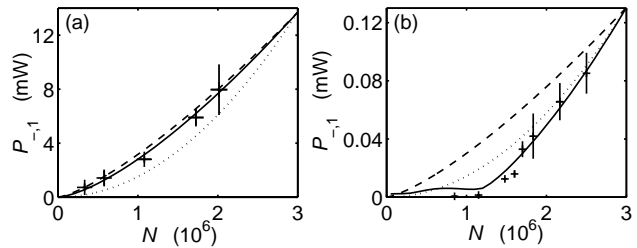}}}\caption{
			(a) Measured height of the first superradiant peak as a function of atom number $N$. The cavity is operated at high finesse
			(good-cavity limit), the pump laser is set to $P_+=0.5~$W and $\lambda=796.1~$nm. The solid line shows a simulation with no
			free parameters. To emphasize the $N$ dependence we also plot curves going like $N^2$ (dotted line) and $N^{4/3}$ (dashed
			line). Those dependences are expected in the superradiant and in the good-cavity limit, respectively.
			(b) Same as (a) but the cavity is now operated at low finesse (superradiant limit), the pump laser is set to $P_+=70~$mW. To
			compensate for the loss in cooperativity the laser is tuned closer to resonance at $\lambda=795.3~$nm. The amount of light in
			the reverse mode due to mirror backscattering in steady-state without atoms is $P_{-,s}=1.8\times 10^{-4}P_+$.}
		\label{Fig3}
	\end{figure}

Clear features that allow us to distinguish between the good-cavity and the superradiant regime are the dependences of the superradiant peak height on the pump power and on atom number: $P_{-,1}\propto N^2P_+$ for the superradiant case and $P_{-,1}\propto N^{4/3}P_+^{1/3}$ for good cavities. The different $N$ dependences are demonstrated in Figs.~\ref{Fig3}(a,b). Interestingly, the superradiant limit exhibits a threshold-like behavior. It is due to backscattering at the mirror surfaces. This phenomenon, which also accounts for the light being in the reverse mode below the threshold, is discussed below. 


Differently from earlier SR experiments \cite{Inouye99,Yoshikawa05} we observe not only a single light burst, but a train of self-similar pulses reminiscent to nonlinear ringing, which demonstrates how the cavity preserves the coherence against diffusion in momentum space induced by the collective coupling. The cavity is however also able to neutralize diffusion due to thermal motion. In fact, we observe SRS for temperatures of the atomic cloud as high as $40~\mu$K (which is 2 orders of magnitude larger than previously reported \cite{Yoshikawa05}). We studied the regime of high temperatures, $\sigma_v\gg\kappa,\omega_r$, earlier \cite{Kruse03b}, but we are now able to study the CARL dependence on temperature down to the level of quantum degeneracy.

\begin{figure}[ht]
		\centerline{\scalebox{0.48}{\includegraphics{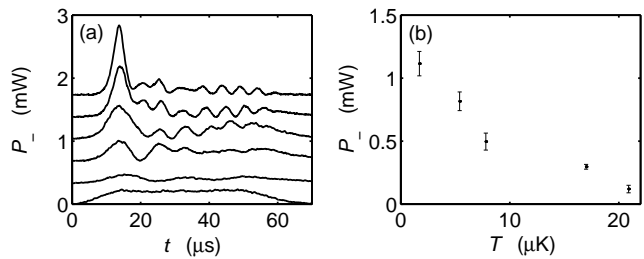}}}\caption{
			(a) Measured time-evolutions of the reverse mode at temperatures between $T=2~\mu$K (upper curve) and $40~\mu$K (lower curve).
			For clarity the curves are vertically shifted from one another by $0.35~$mW. The atom number is the same for all data points,
			$N=10^6$. The pump power is $P_+=1~$W. The reverse power observed at high temperatures is there even without atoms and is
			caused by mirror backscattering.
			(b) Measured SRS peak height as a function of temperature.}
		\label{Fig4}
	\end{figure}

We experimentally tune the temperature via forced evaporation ensuring a fixed final atom number of $N=10^6$ for all measurements. Fig.~\ref{Fig4}(a) show how the temperature affects the evolution of the reverse power. At low temperatures, strong SR bursts are observed followed by a large number of revivals. As the temperature rises, the revivals gradually fade out, until at $T\simeq20~\mu$K only a single burst survives. The amount of power scattered into the reverse mode rapidly decreases at higher temperatures, as seen in Fig.~\ref{Fig4}(b).

At even higher temperature beyond $40~\mu$K we observe that the reverse power, $P_-$, approaches a stationary value within the time $\tau_{bw}$ without discernible oscillations. In fact, in this case the evolution of the reverse power is the same as for the empty cavity. It is due to residual backscattering from impurities at the mirror surfaces. The large amount of mirror backscattering, which adopts values between $0$ and $1\%$ \cite{Note2}, suggests that it plays a major role in the dynamics. This is true indeed for small atom numbers and in the very beginning of the light pulse. But as soon as the reverse mode starts to build up, the collective dynamics takes over completely, and the rise time of the reverse mode intensity is much shorter than the cavity build-up time. The height of the SR peaks turns out to be quite insensitive to the amount of mirror backscattering, and at longer times the average value of the light intensity is smaller than in the absence of atoms, i.e.~the collective atomic motion suppresses mirror backscattering.


In conclusion, our observation of superradiance in the good-cavity regime of the CARL experimentally demonstrates the intrinsic link between both phenomena. Together with the earlier observation of CARL with atomic clouds as hot as several $100~\mu$K \cite{Kruse03b} this proves that the gain process underlying both, SRS and CARL, is not based on quantum statistics, but on cooperativity \cite{Moore01}. These results clarify the intricate relationship between CARL and superradiance in a regime, where the coupling between radiative and matter-wave modes is completely coherent. This is also the regime for which the prospect of a robust quantum entanglement of the modes has been pointed out \cite{Moore99}. 

Although the experiment reported here is the first one to study the interaction of Bose-Einstein condensates with macroscopic cavities, the quantum degeneracy plays no role in the observed effects. Future studies however, may focus on the role of quantum statistics and interatomic interaction \cite{Horak00}.

Another challenge would be to reach the so called quantum limit. This limit is distinguished from the semiclassical limit by the fact that the gain bandwidth is so small, $\Delta\omega_G\ll\omega_r$, that only adjacent momentum states of the atomic motion are coupled. This case (provided the temperature is very low) results in a train of self-similar superradiant pulses \cite{Piovella01b}. For us this regime could be reached for increased finesse or by increasing $\omega_r$, e.g.~by tuning the pump laser to an atomic resonance at a much higher frequency. To treat this regime the use of quantized atomic motion in the CARL equations is compulsory \cite{Piovella01b}. 


This work has been supported by the Deutsche Forschungsgemeinschaft (DFG) under Contract No. Co~229/3-1.

\end{document}